\documentclass[aps,onecolumn,preprint,superscriptaddress,nofootinbib,floats]{revtex4}
\usepackage{amsmath,amssymb,color,mathrsfs, graphicx,verbatim,epsfig, bbm, wasysym,slashed}
\usepackage[hyperfootnotes=false]{hyperref}
\usepackage{slashed}
\allowdisplaybreaks

\def\lsim{\mathrel{\rlap{\lower4pt\hbox{\hskip1pt$\sim$}}
    \raise1pt\hbox{$<$}}}
\def\gsim{\mathrel{\rlap{\lower4pt\hbox{\hskip1pt$\sim$}}
    \raise1pt\hbox{$>$}}}

\newcommand{\be}{\begin{eqnarray}}
\newcommand{\ee}{\end{eqnarray}}
\newcommand{\cL}{{\cal L}}

\newcommand{\met}{{\slashed E_T}}

\def\addresses#1#2{\hbox to \hsize{\@tablebox{#1}\hfil\@tablebox{#2}}}
\def\@tablebox#1{\vtop{\hsize=5in \begin{flushleft} #1 \end{flushleft}}}

\def\beq{\begin{equation}}
\def\eeq{\end{equation}}
\def\bit{\begin{itemize}}
\def\eit{\end{itemize}}
\def\beqa{\begin{eqnarray}}
\def\eeqa{\end{eqnarray}}

\def\MadGraph{{\tt MadGraph}}
\def\MadGraph5{{\tt MadGraph5}}

\def\stop{\tilde t}

\def\mstop{m_{\stop}}

\def\met{{\slashed E}_T}

\begin{document}

\baselineskip 0.6cm

\begin{titlepage}

\thispagestyle{empty}

\begin{flushright}
\end{flushright}

\begin{center}

\vskip 2cm

{\Large \bf Stealth Stops and Spin Correlation: A Snowmass White Paper }

\vskip 1.0cm
{\large  Zhenyu Han$^1$ and Andrey Katz$^2$}
\vskip 0.4cm
{\it $^1$ Institute of Theoretical Science, University of Oregon, Eugene, OR 97403}\\ 
{\it $^2$ Center for the Fundamental Laws of Nature, Jefferson Physical Laboratory,\\ 
Harvard University, Cambridge, MA 02138} 
\vskip 1.2cm

\end{center}

\noindent
Stops with the mass nearly degenerate with the top mass, decaying into tops and soft neutralinos, are usually dubbed ``stealth stops''. Their kinematics looks very similar to that of the standard $t \bar t$ events, which leads to events with little or no excess of missing transverse energy. This complicates the probing of this region of the stop parameter space by hadron colliders, rendering the application of standard searching techniques challenging. In this Snowmass white paper we reanalyze the spin correlation approach to the search of stealth stops, focusing on the feasibility of this search at the 14~TeV LHC. We find, while the statistical limitations significantly shrink compared to the low-luminosity 8~TeV run, the systematic PDF uncertainties pose the main obstacle. We show that the current understanding of PDFs probably does not allow us to talk about top and stop discrimination via spin correlation in the inclusive sample. On the other hand the systematic uncertainties significantly shrink if only events with low center of mass energy are considered, rendering the search in this region feasible.

\end{titlepage}

\setcounter{page}{1}

\section{Introduction}
\label{sec:intro}
Current experimental searches of ATLAS and CMS strongly constrain colored SUSY-particle production below the TeV scale, resulting bounds on gluinos or mass degenerate squarks well above 1 TeV~\cite{ATLAS-CONF-2013-047}. However there still exist a few important examples of new physics,  in which new particles are allowed below the TeV scale. In spite of the enormous production cross-section, the signal is deeply buried either in the QCD or the $t \bar t $ background. While certain SUSY particle spectra with baryon-number violation (e.g. light up-type squarks decaying into a pair of  ``anonymous jets'') can be a good example for the former, the most well-known example for the latter is perhaps the ``stealth stop''.  

\begin{figure}[t]
\centering
\includegraphics[width=0.99\textwidth]{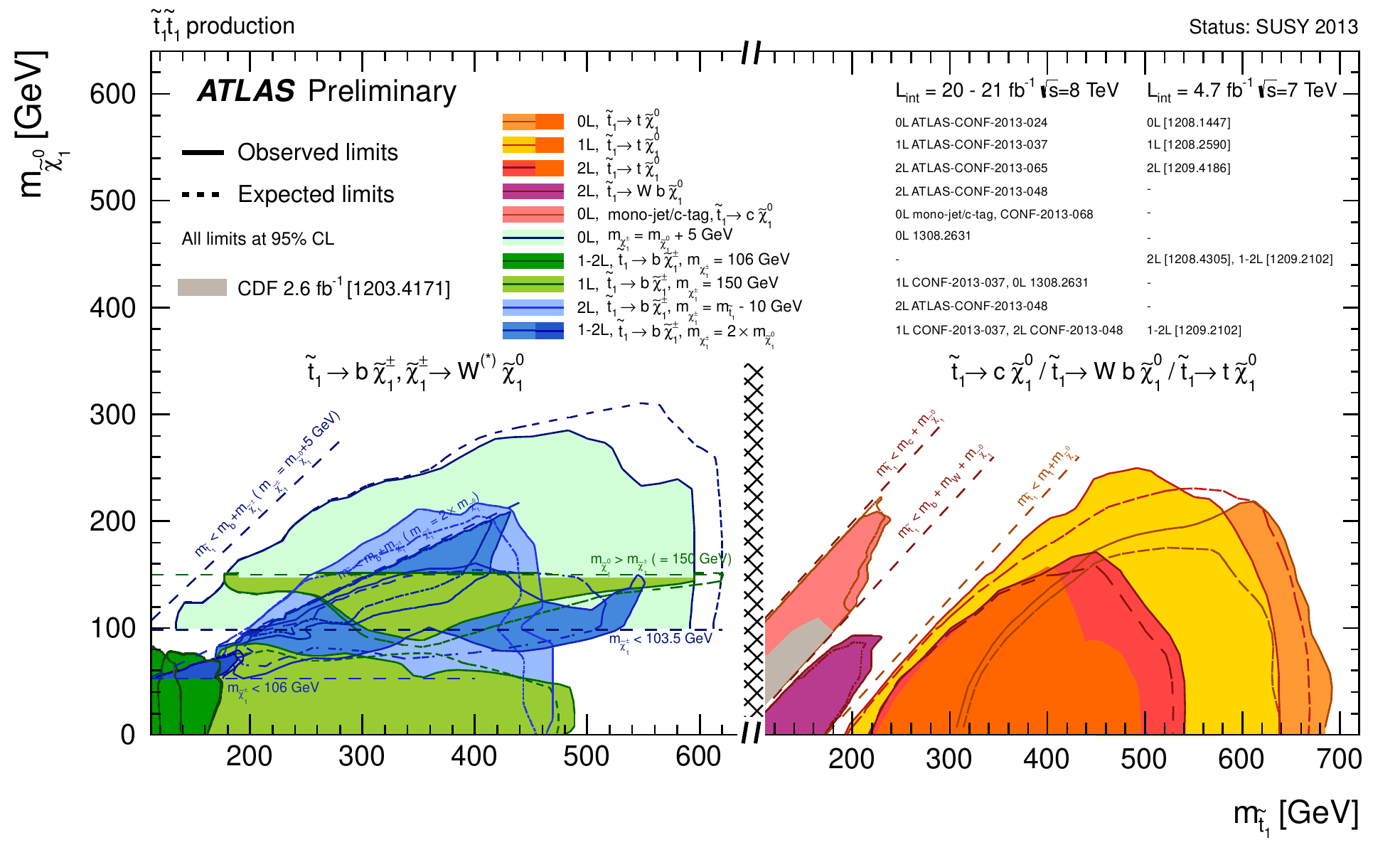}
\caption{ATLAS Constraints on stops decaying into chargino (on the left) and to the LSP neutralino (on the right). 
The plot manifestly shows that the stealth regime: $175~{\rm GeV} < \mstop < 200~{\rm Gev}$ are not excluded.   }
\label{fig:exclusions}
\end{figure}

An R-parity conserving NLSP stop, which was originally motivated in ``natural SUSY''~\cite{Dimopoulos:1995mi,Cohen:1996vb,Kats:2011qh,Brust:2011tb,Papucci:2011wy}, is often called ``stealth'', when its mass is nearly degenerate with the top mass, and the LSP neutralino is almost massless. In this case the stop decays into a top and the LSP. Due to the nearly degenerate top and stop masses, the LSP carries a very low momentum in the mother-particle rest frame. Due to its small mass, boosting it to the lab frame does not change the situation. This results in stealth $\tilde t \tilde t^*$ events with the LSP carrying very little missing energy, on top of the contributions of the neutrinos from leptonic top decays. As a result the kinematics of $\tilde t \tilde t^* $ events closely resemble the kinematics of $t\bar t$ events, with almost no excessive $m_T$ or $m_{T2}$, rendering the corresponding searches inefficient. This situation is illustrated in 
Fig.~\ref{fig:exclusions}, which shows the summary of the constraints on an NLSP stop decaying into the LSP neutralino,  based on ATLAS searches with $\met$~\cite{Aad:2012ywa}, $m_T$~\cite{Aad:2012xqa} and $m_{T2}$~\cite{Aad:2012uu} cuts. The region around $\mstop = 175$~GeV is not excluded, and the searches, which are based on missing transverse energy techniques, are all inefficient in this part of the parameter space. 

Note that unlike in the compressed SUSY case, we do not get any significant contribution of $\met$ in the events when stealth stops recoil against hard ISR jets~\cite{Fan:2012jf}. This happens because in the compressed spectra, the LSP is massive and a large boost will result in a large momentum for the LSP and thus large $\met$. On the other hand, for a nearly massless LSP as in the stealth regime, the momentum remains small even when boosted.

\begin{figure}[t]
\centering
\includegraphics[width=0.7\textwidth]{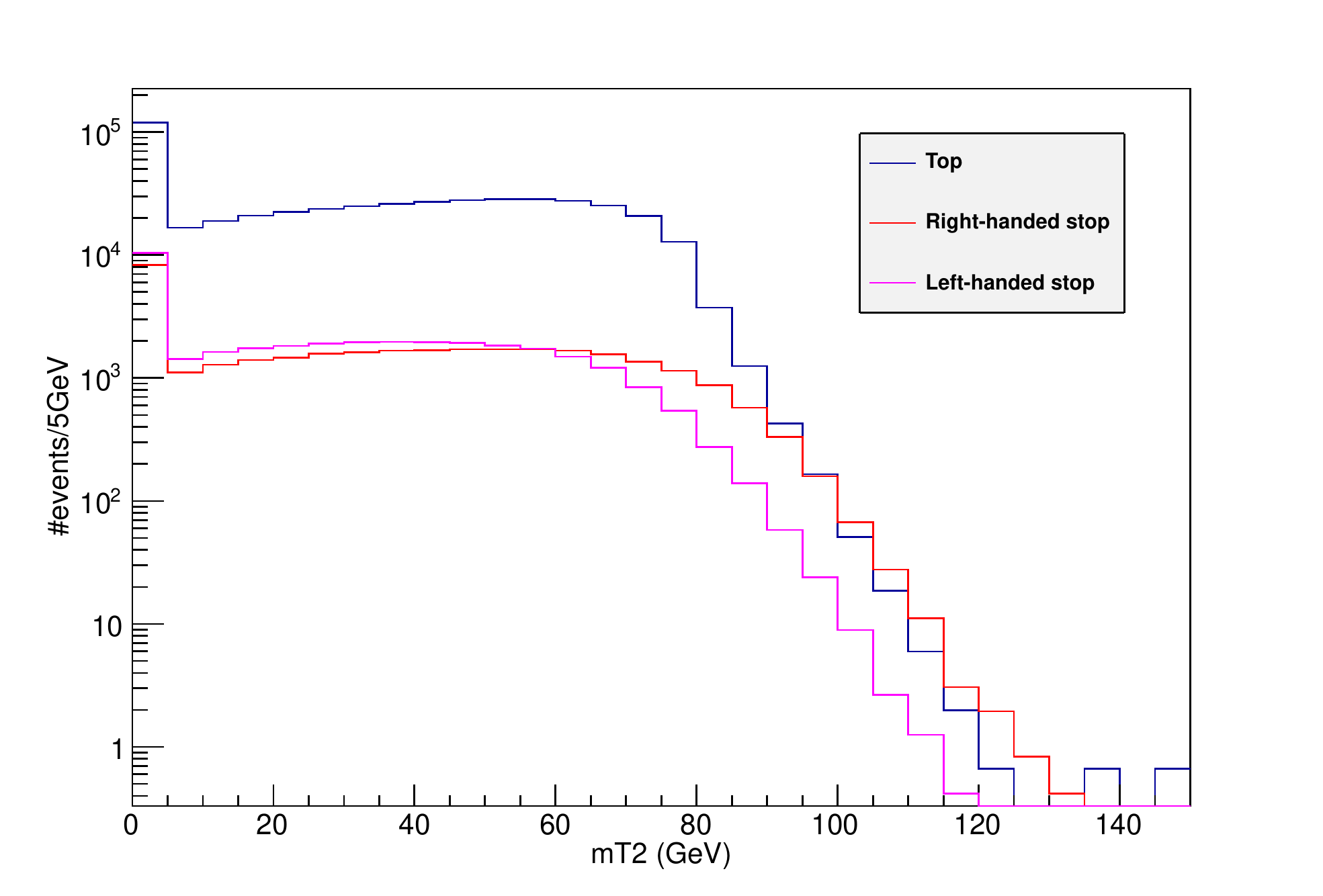}
\caption{$m_{T2}$ distribution for different stop chiralities decaying into bino-like neutralino. RH stops are much more perspective for this search than its LH counterpart.}
\label{fig:mt2}
\end{figure}

Two different techniques have been proposed in the literature to tackle the stealth stop spectrum. One of them, considered in Ref.~\cite{Kilic:2012kw}, takes advantage of the clean dileptonic decay channel, in which a small number of events have a larger $m_{T2}$. This is because, when the stop mass becomes very close to the top mass, one has a small number of events where the stop decays through an off-shell top (although two-body decay $\stop \to t \tilde \chi^0$ is kinematically allowed) and a relatively more energetic LSP, leaving a sufficient number of events with large $m_{T2}$. The effect strongly depends on the chirality of the stop and the nature of the LSP (bino vs higgsino). In particular, the authors noticed that right-handed (RH) stops decaying into bino-like neutralinos or left-handed (LH) stops decaying into 
higgsino-like neutralinos tend to have a larger $m_{T2}$ than, for example LH stops decaying into bino, or RH stops decaying into higgsinos. We show this feature explicitly in Fig.~\ref{fig:mt2}

Another technique, which we proposed in~\cite{Han:2012fw}, takes advantage of the spin correlation in $t \bar t$ events. Top-quarks are unique in the sense that they usually decay before they hadronize, and therefore lots of their features can be directly measured from the kinematic distributions of their decay products. Notably, one of these important features is the spin correlation. Since top-quarks are fermions, spins of the quarks are correlated with one another, a feature which can be measured from the angular distribution of its decay products (see Ref.~\cite{Parke:2010ud} for a review, and Ref.~\cite{Baumgart:2011wk} for other new physics applications). These searches are mostly done in the dileptonic channel, although it may also be carried out in the less clean, but much more abundant semileptonic channel, which yields a comparable sensitivity. 

Unlike tops, stops are scalar particles, and carry no spin correlation information. Consequently, tops from stop decays are spin-uncorrelated. This is in particular true for stealth stops, which yield $\met $ distributions similar to 
those of tops, but are expected to have different angular distributions for its decay products. In this note we take advantage of this difference between tops and stops events and apply the technique proposed in Ref.~\cite{Han:2012fw} study the discovery potential of the 14 TeV LHC. The technique is largely insensitive to
the chirality of the decaying stop or the nature of the LSP higgsino, and therefore this technique and the $m_{T2}$ technique of ~\cite{Kilic:2012kw} should be considered complementary for covering  the entire parameter space of the stealth stops. We also study the systematic uncertainties, especially the PDF uncertainties, associated with this technique. 

Our white paper is organized as follows. In Sec.~\ref{sec:review} we review the theoretical background and the current
experimental status of spin correlation measurement in $t \bar t$ events. In Sec.~\ref{sec:analysis}, we perform our analysis and discuss PDF uncertainties. Finally, we conclude in Sec.~\ref{sec:conclusions}.

\section{Techniques}
\label{sec:review}
Top pairs production at the LHC at low invariant mass is dominated by fusion of same-helicity gluons, such that the 
final state $t \bar t$ pairs are either both RH or both LH~\cite{Mahlon:2010gw}. On the other hand, high invariant mass
processes are largely dominated  by opposite-helicity gluons and DY, leading to opposite helicity $t \bar t$.  

This fact, that the tops have like-helicity in low-mass events are reflected in various angular distributions of $t \bar t$ decay products, most noticeably the azimuthal angle difference between the two leptons in dileptonic events, $\Delta \phi_{ll}$. We illustrate this point in Fig.~\ref{fig:dphill} (plots from Ref.~\cite{Mahlon:2010gw}). This dependence was used LHC searches to demonstrate the existence of spin correlation in $t \bar t$
samples with more than $5 \sigma$ significance \cite{ATLAS:2012ao, CMS-PAS-TOP-12-004}. 

\begin{figure}
\begin{center}
\begin{tabular}{cc}
\includegraphics[width=0.5\textwidth]{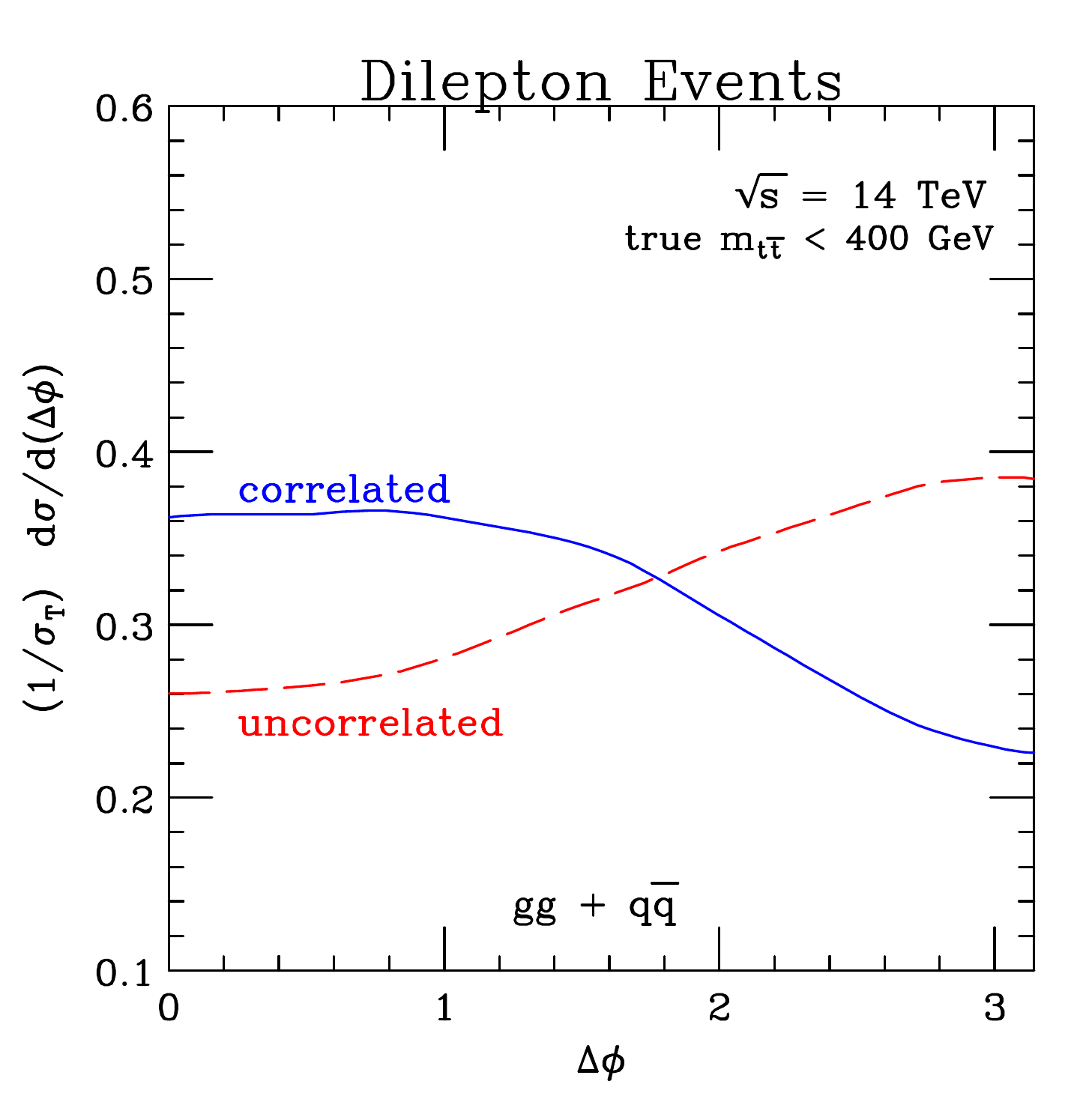}
&\includegraphics[width=0.5\textwidth]{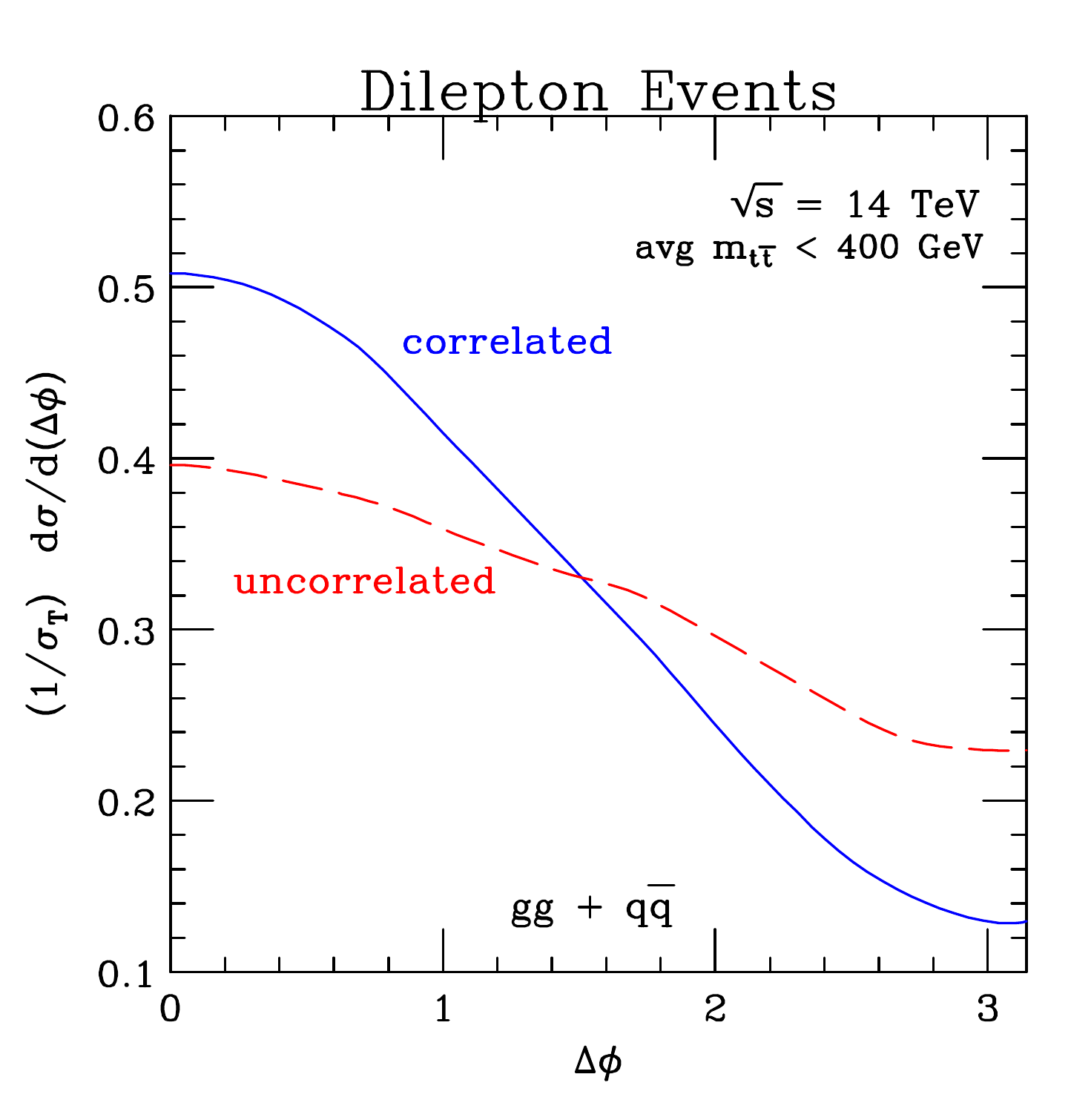}
\end{tabular}
\caption{The differential distribution of $\Delta\phi$}. Left: a cut on the true $m_{t\bar t}$ is imposed; Right: a cut on the reconstructed $m_{t\bar t}$, averaged over solutions is used. 
\label{fig:dphill}
\end{center}
\end{figure} 

In our case the problem is more challenging than just establishing spin correlation in $t \bar t$ sample. We are trying to distinguish
between a \emph{pure $t \bar t$ sample} and a sample contaminated by stop events at the level of $O(10\%)$. As discussed in the introduction, the latter mostly behave as spin-uncorrelated tops. In order to exploit all available spin correlation information, we proceed with the full matrix element method proposed in Ref.~\cite{Melnikov:2011ai}. 

Following Ref.~\cite{Melnikov:2011ai}, we define a probability distribution for both the correlated and the uncorrelated hypotheses 
\beq \label{eq:PDist}
P _H = {\cal N}^{-1}_H \sum_{ij} \sum_a J_a f_i^{(a)} f_j^{(a)} \left| {\cal M}^{ij}_H (p_{obs}, p_\nu^{(a)},
 p_{\bar nu}^{(a)})\right|^2~,
\eeq
where $H$ stands for the correlated or the uncorrelated hypothesis, $f_i$ are parton distribution functions of the incoming partons and ${\cal M}$
is a leading order matrix element. A-priori we do not know the neutrino momenta, and therefore $J_a$ is a Jacobian obtained when integrating over the neutrino momenta. The uncorrelated LO matrix element describes a spherically symmetric decay of the top quark 
into a $b$-quark and $W$-boson. We take all the expressions for the LO matrix elements, both spin-correlated and spin-uncorrelated, from Ref.~\cite{Mahlon:2010gw}. 

We further define for each event a likelihood for the event to be a correlated top pair,
\beq
{\cal R} = \frac{P_{corr}}{P_{corr} + P_{uncorr}}~.
\eeq 
Finally, given the likelihood distribution, for a given event sample we define a log-likelihood ratio 
\begin{equation}
L = 2\ln \frac{\cL_t}{\cL_{\bar t}}, \ \  \ {\rm where } \ \ \ \cL_K = \prod_i^N \rho_K ({\cal R}_i)
\end{equation}
where $\rho_K $ is a probability density read from the likelihood distributions; the product is over the $N$ events in the sample. Example $L$ distributions for the pure top hypothesis and the top+stop hypothesis are given in Fig.~\ref{fig:L-all}. If the two distributions are well seperated, we will be able to distinguish the two hypothesis.

\section{Analysis for $\sqrt{s} = 14$~TeV LHC and discussion of systematic uncertainties}
\label{sec:analysis}
In our LHC14 analysis we closely follow the steps in Ref~\cite{Han:2012fw}. We generate both $t \bar t$ and $\stop \stop^*$ events with 
{\tt MadGraph~5} \cite{Alwall:2011uj}. For the signal, we simulate both pure left-handed and pure right-handed stops with a mass $\mstop = 200$~GeV decaying into a massless (bino-like) neutralino. As it can be seen from Fig.~\ref{fig:exclusions}, $\mstop = 200$~GeV is right on the edge of the standard techniques reach, and as we go deeper into the stealth regime, reducing the stop mass, our sensitivity will in general improve, since the stop production cross section will grow. As in~\cite{Han:2012fw} we find very little difference between the stops with different chiralities.   
 
We further shower the $t \bar t$ and $\stop \stop^*$ events with {\tt Pythia 6}~\cite{pythia6} and cluster them with {\tt FastJet~3}~\cite{fastjet}. To mimic the detector effects we process all hadronic particles through a perfect $0.1\times 0.1$ grid hadronic calorimeter in $\eta - \phi$ space. For our analysis we consider events with precisely two isolated leptons with $p_T > 20$~GeV and $|\eta| < 2.5$, at least two jets with $p_T > 50$~GeV and $|\eta| < 2.5$, at least one of them should be b-tagged. We assume a 60\% b-tagging rate, regardless of the $p_T$ or $\eta$ of the b-jet. If the leptons have opposite signs with the same flavor, we demand $\met > 40$~GeV and veto events with the dilepton invariant mass close to the Z mass, namely, we demand $|m_{ll}-m_Z| > 10$~GeV. 

In the selected events we assume that the entire missing energy in the event is carried by two neutrinos. Using four mass shell conditions (two for the
$W$s and two for the tops) we can reconstruct the momenta of both neutrinos up to discrete degeneracies. One can have zero, two or four real solutions. Events without real solutions are discarded, and the total acceptance is 
17.1\% for top events and 17.5\% for left handed stop events\footnote{Compare to 16.8\% and 15.6\% respectively at LHC 8.}. We take all these solutions into account in calculation of the log likelihood ratio as explained in Sec.~\ref{sec:review}.

\begin{figure}[t]
\centering
\includegraphics[width=0.6\textwidth]{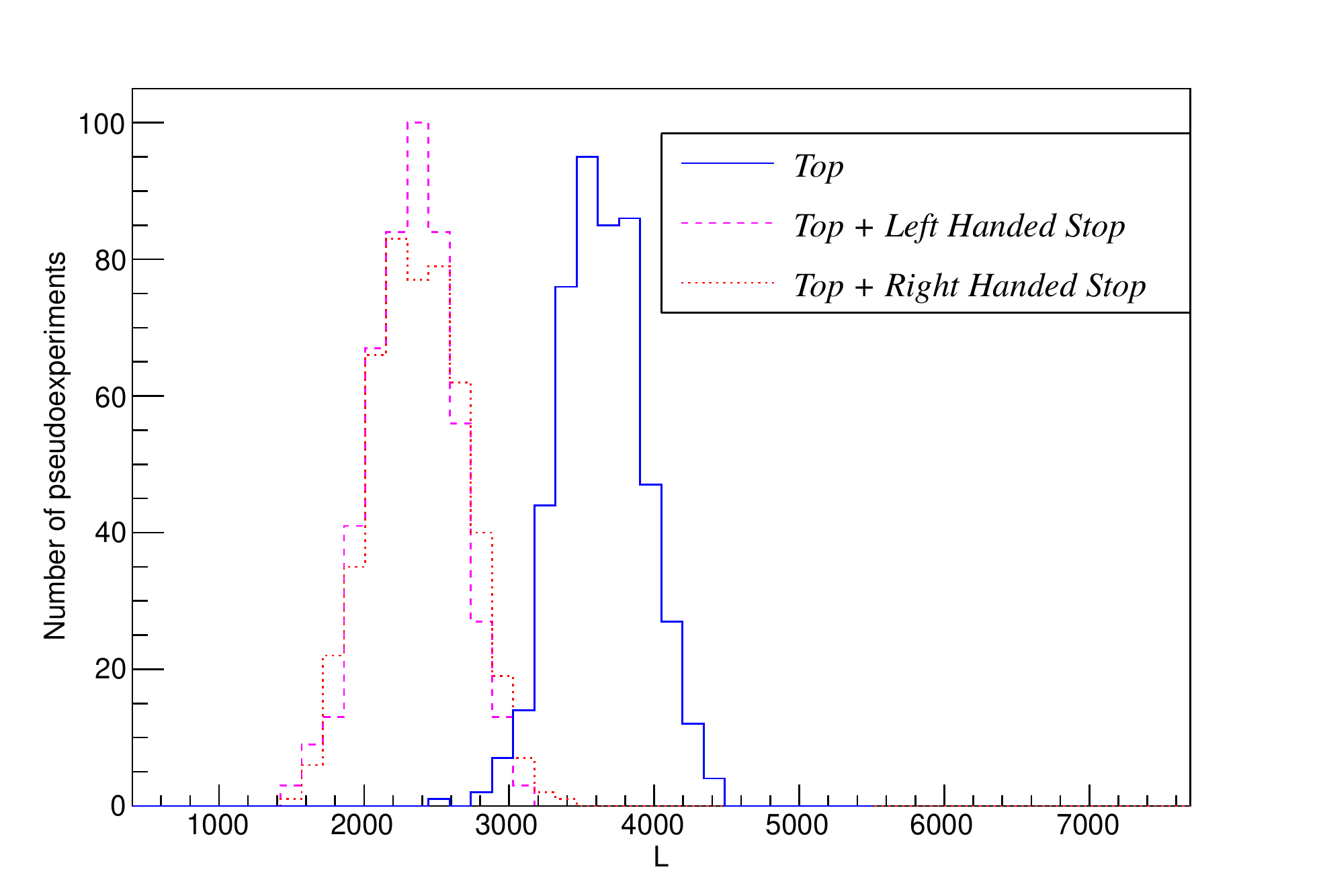}
\caption{The log likelihood ratio L. Each point on the curves corresponds to a pseudo-experiments at LHC~14 with 100 fb${}^{-1}$ data. Jet level results with the cteq6l1 PDFs are shown. We assume $\mstop = 200$~GeV, for lower stop masses we anticipate even better results due to larger $\stop \stop^*$ 
cross-sections}
\label{fig:L-all}
\end{figure} 

It was found in~\cite{Melnikov:2011ai} that NLO matrix element corrections are almost negligible at the LHC, and therefore we neglect these corrections here. We present the log-likelihood ratio for the $t \bar t$ and $\tilde t \tilde t^*+ t \bar t$ samples in Fig.~\ref{fig:L-all}. The ratio between the number of $t \bar t$ events and the number of $\tilde t \tilde t^*+ t \bar t$ events is 14. We see that at LHC 14 for integrated luminosity of $\cL  = 100$~fb$^{-1}$ one should have an excellent separation between tops and stops,  $\sim5\sigma$, mainly because we face almost no statistical limitation. We also see clearly that there is little difference in the log-likelihood distributions for the LH and the RH stops, showing that the chirality of the stealth stop plays no role in our procedure.    

However, statistics is not the only limitation for this measurement. Although the systematics of the spin-correlation matrix element
is claimed to be well-understood, we also have uncertainties coming from the PDFs. We did not study these systematic 
uncertainties in our original paper, and we are addressing this issue in the current note. First we check if changing the LO PDFs 
(we use the central values of {\tt cteq6l1} as our nominal LO PDF) to the 
NLO PDF changes the relevant kinematic distributions. Performing a similar analysis with {\tt cteq6m} NLO PDFs, we notice that 
log-likelihood distributions significantly change compared to the LO PDFs, signaling that NLO corrections can be an important effect in this analysis. 

\begin{figure}[t]
\centering
\includegraphics[width=0.6\textwidth]{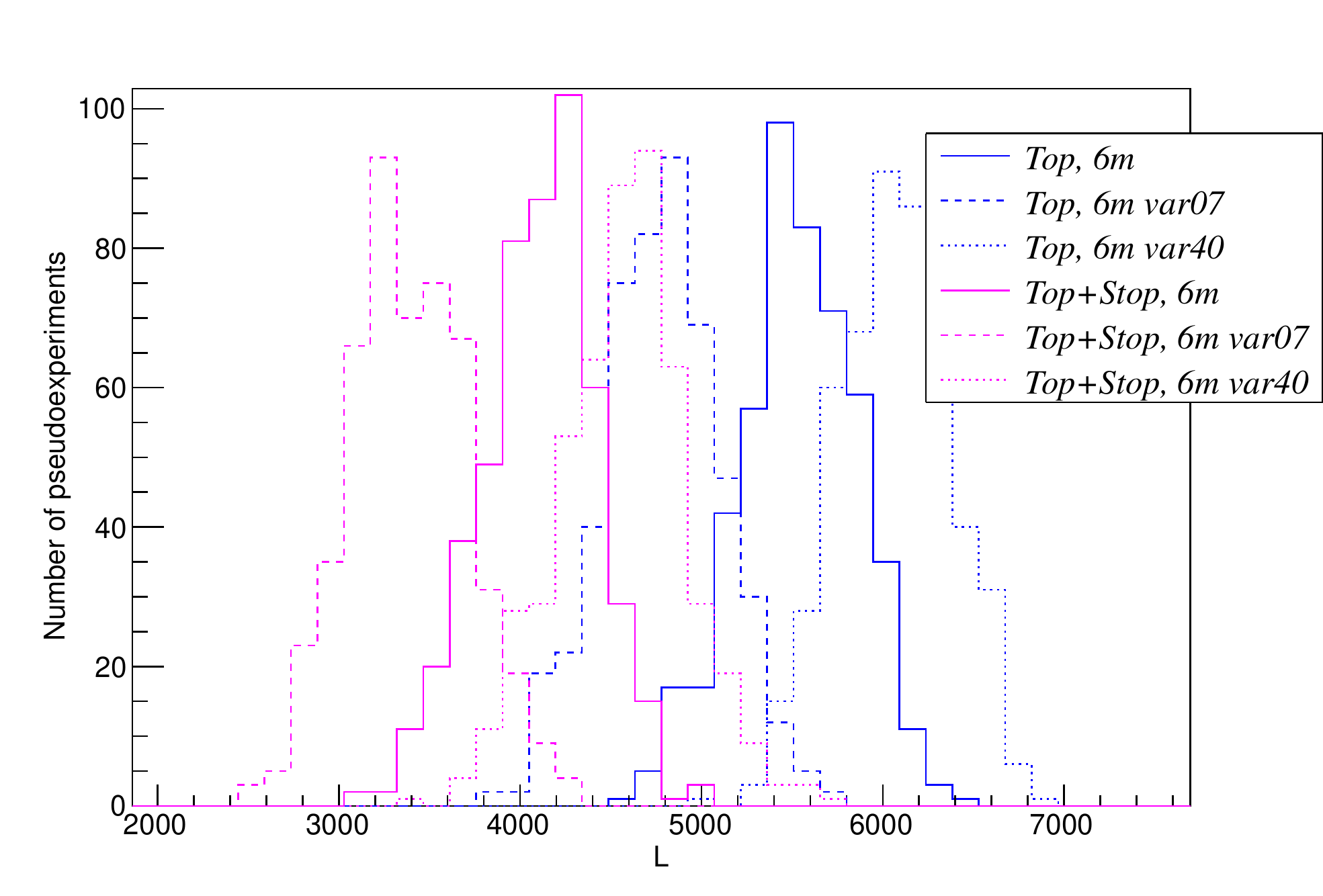}
\caption{The log likelihood ratio L for cteq6m variations.}
\label{fig:L-all-6m-vars}
\end{figure}

We further proceed with the NLO PDF analysis, checking for the potential systematic errors, potentially induced by the PDF 
uncertainties. We vary the {\tt cteq6m} PDFs within $2\sigma $ uncertainties. Forty different 
 variations are coded in {\tt MadGraph 5},  
we randomly choose  seven different variations from the central values, each of which represents an allowed deviation from the 
central values within $2\sigma$. We find that these uncertainties are appreciable and can invalidate the entire results of our 
previous study. We plot the log-likelihood of two of these allowed variations on Fig~\ref{fig:L-all-6m-vars}. Clearly, 
we will not be able to distinguish between the pure tops and the mixed samples, since the deviations from 
the predicted SM central values can be easily swallowed into PDF uncertainties.  Therefore, if the entire $t \bar t$ sample 
is used in a search like this at LHC~14, one would need to improve the understanding of PDFs significantly, before any 
definitive conclusion about either the discovery or the exclusion of the stealth stops is made.

The reason for these uncertainties is relatively easy to understand.  The spin correlation 
in quark-quark and gluon-gluon events are manifestly different. Therefore we have to use different matrix elements 
in our probability  distribution~\eqref{eq:PDist}. Our uncertainties in $q\bar q$ vs $gg$ initial states can directly 
affect the interpretation of this measurement.

Given this somewhat frustrating result, it would be interesting to explore, where these uncertainties come from, and whether there
are any parts of the parameter space where they can at least be ameliorated. As discussed in Ref.~\cite{Mahlon:2010gw}, it helps to separate the events to high and low invariant mass regions, where gluon fusion produces top quarks with different spin correlations. For further studies we divide all our events into two bins: 
low effective mass events and high effective mass events. We define the effective mass of the event as 
\beq
M_{eff} = p_T(l_1) + p_T(l_2) + \sum_{i} p_T(j_i) +\met.
\eeq  
Interestingly, the spin correlation in the $t\bar t$ sample is stronger in the low effective mass events, making our results less sensitive to the PDF uncertainties. This can be first noticed in the difference in azimuthal angles for the two charged leptons, 
$\Delta \phi_{ll}$, which is, as explained in Sec.~\ref{sec:review}, one of  the key variables in our spin correlation analysis. The distributions of the azimuthal angle difference for all events and the low effective
mass events are shown in Fig.~\ref{fig:dphill}. It is easy to see that even if we just use an azimuthal angle as a sole discriminator, we should expect much smaller uncertainties and stronger effect in the low $M_{eff}$ events, than in the inclusive sample.

\begin{figure}
\begin{center}
\begin{tabular}{cc}
\includegraphics[width=0.45\textwidth]{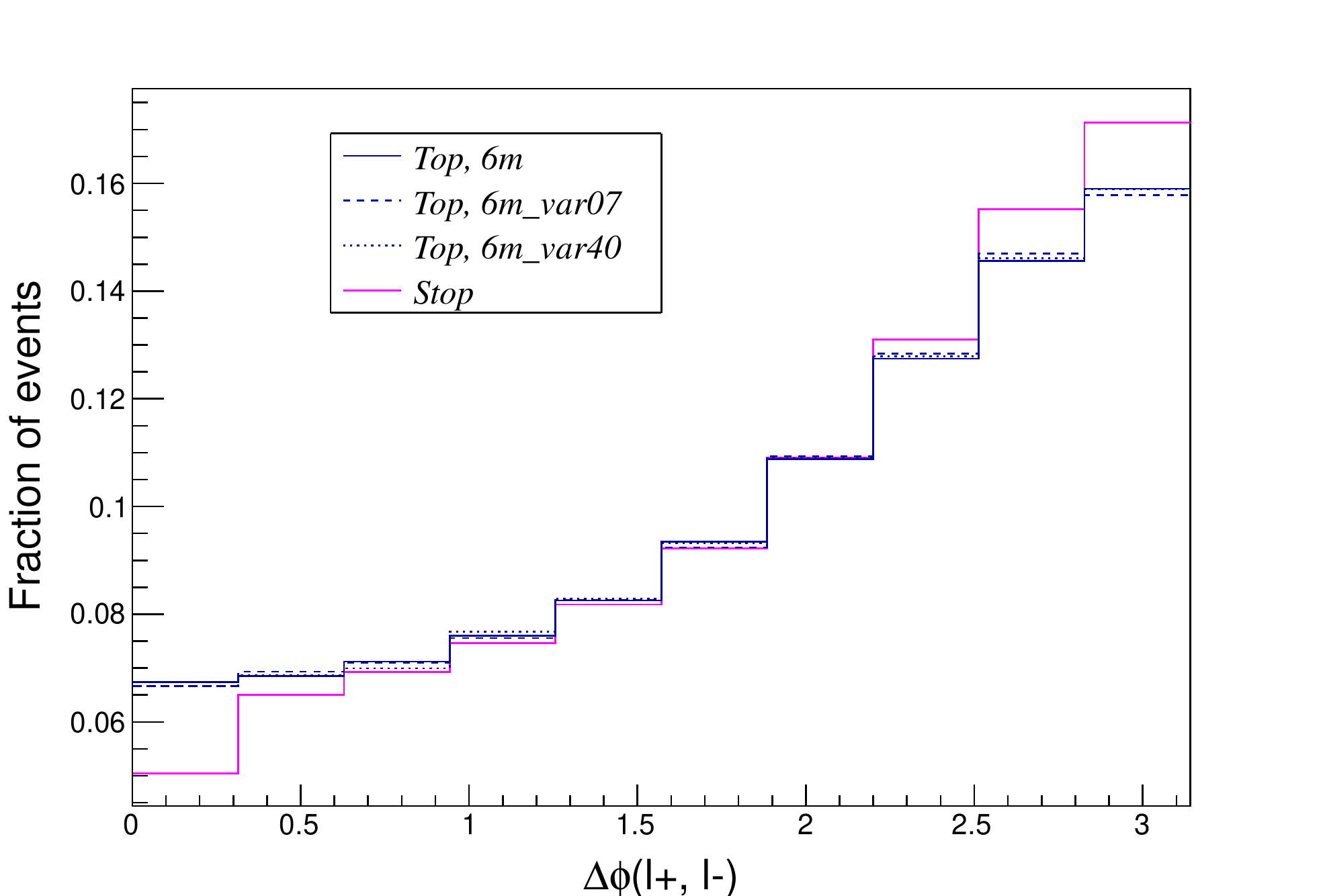}
&\includegraphics[width=0.45\textwidth]{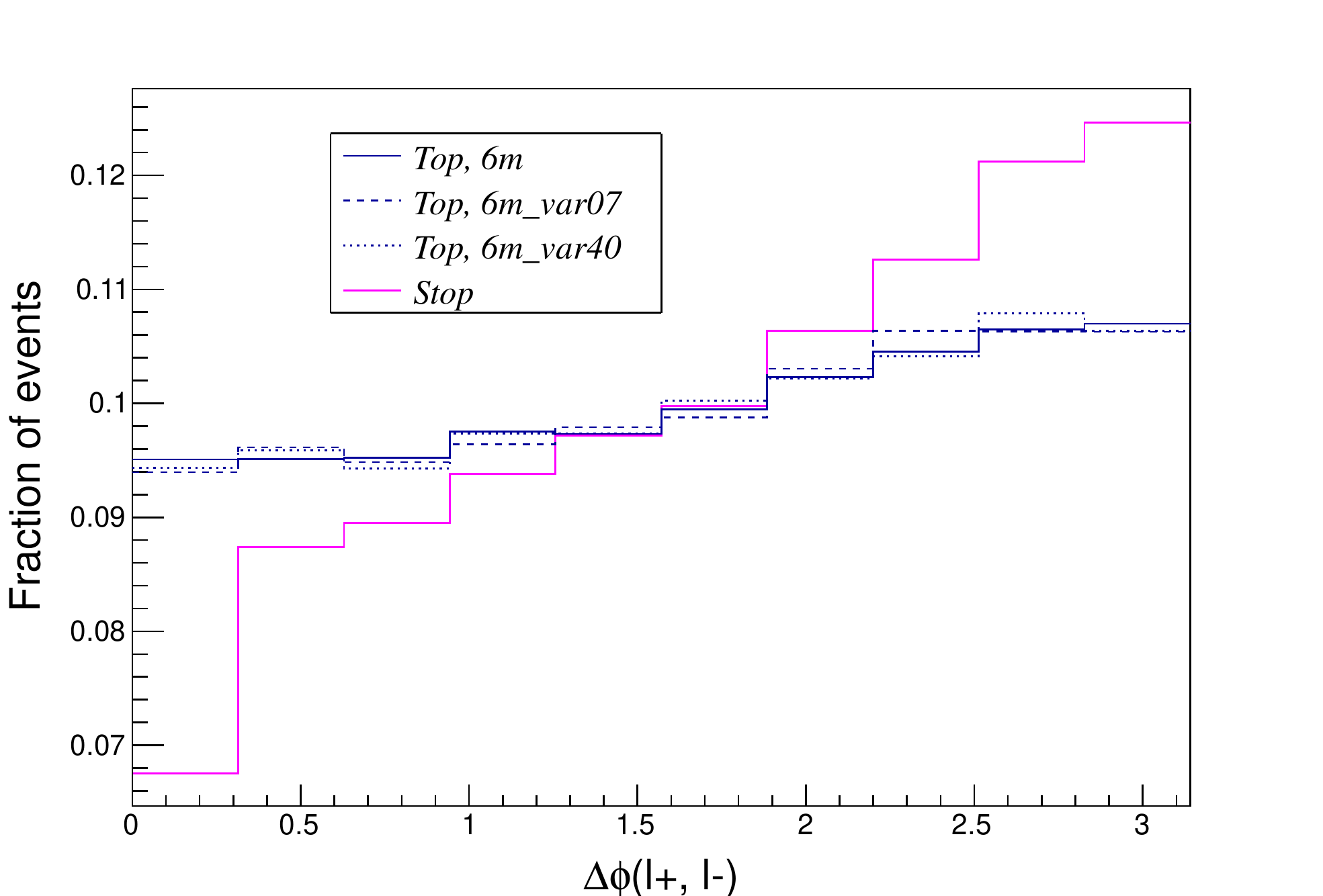}
\\(a) &(b)
\end{tabular}
\caption{$\Delta\phi(l^+,l^-)$ distribution. (a): all events; (b): events with $m_{\text{eff}}<400$ GeV.}
\label{fig:dphill}
\end{center}
\end{figure} 

\begin{figure}[hbt]
\centering
\includegraphics[width=0.6\textwidth]{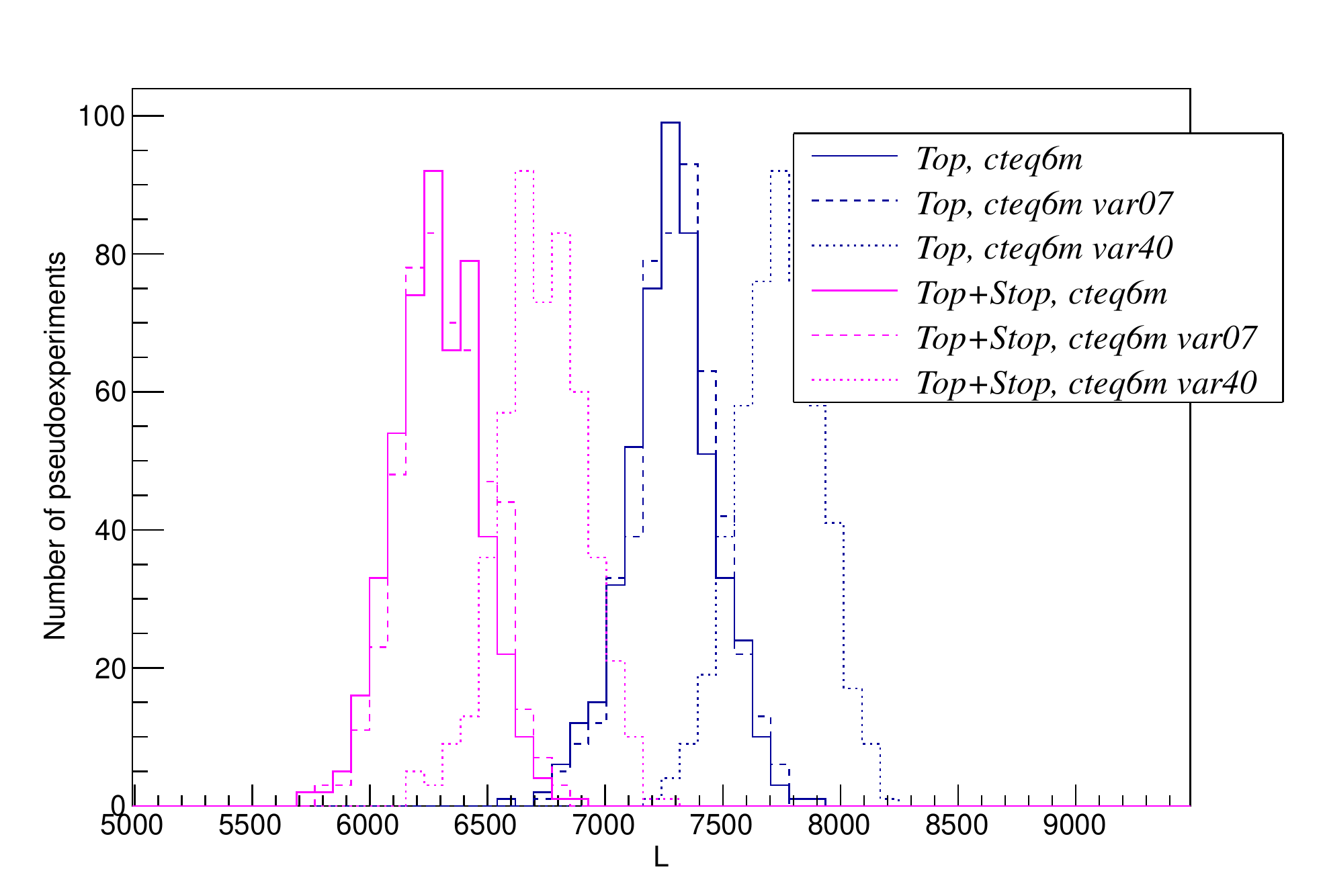}
\caption{The log likelihood ratio L for cteq6m variations with $m_{\text{eff}} < 400$ GeV.}
\label{fig:L-all-6m-vars-lowmeff}
\end{figure}

We further construct the log-likelihood ratios for the signal and background event samples with low $M_{eff}$. The results are presented in  Fig.~\ref{fig:L-all-6m-vars-lowmeff}. We clearly see that even for the worst possible assumptions about PDF uncertainties 
the signal+background distribution is very well separated from the background only hypothesis. In this short note, we do not attempt to estimate 
the statistical significance of the method, although the result looks promising, and we expect that with the cut on $M_{eff}$ very good sensitivity can be achieved. 

Note that in this white paper we have always used the leading order matrix element, in both our event simulation (though extra jets are added by Pythia from parton showers) and likelihood calculation. One should obtain better results by using NLO matrix elements and matched samples, together with the NLO PDFs. Also, we did not simulate other effects, such as the contamination from other backgrounds ($t \bar t$ with $\tau$ decays or $Z^*+$jets). These effects were addressed in our original paper and found to be subdominant for LHC~8. We expect them to remain subdominant at LHC~14.

\section{Conclusions and Outlook}
\label{sec:conclusions}
In this short white paper we estimated the feasibility of spin-correlation techniques in searches for stealth stops. We showed that 
LHC~14 with $\cL = 100$~fb$^{-1}$ integrated luminosity will have enough statistics to perform precision measurements of 
$t \bar t$ spin correlation, and if systematic is disregarded one should have $5\sigma $ significance or more for the stealth stops. 

On the other hand we find that systematic uncertainties pose a real challenge to this kind of measurements. Although NLO matrix element uncertainties are probably not a worry, one faces a serious problem from the PDF uncertainties. The 
$t \bar t$ spin correlation is different in $q \bar q$  and $gg$ events, and therefore uncertainties in their distributions translate directly into uncertainties in spin-correlation. In particular, we find that with the current knowledge of PDFs at $\sqrt{s} = 14$~TeV, systematic uncertainties do not allow a good discrimination between tops and tops + stealth stops in the inclusive sample. However, uncertainties become much less pronounced in the low $M_{eff}$ sample, where the 
search has a good potential to succeed. 

We conclude, that probably more work is needed to achieve an acceptable reach to the stealth stops scenario. We hope that our observations, together with improved knowledge of PDFs, will finally allow us to discover or exclude this interesting and challenging regime.


\acknowledgments{We are grateful to Marat Freytsis, Matt Schwartz and Matt Reece for useful discussion. 
ZH was supported in part by DoE grant No.\ DE-FG-02-96ER40969. AK was supported by NSF grant No.\ PHY-0855591.   }



\bibliography{lit}
\bibliographystyle{apsper}

\end{document}